\documentstyle[sprocl]{article}

\newcommand{\nc}{\newcommand}
\nc{\renc}{\renewcommand}

\def\GeV{{\rm\ GeV}}

\def\lsim{\; \raise0.3ex\hbox{$<$\kern-0.75em
      \raise-1.1ex\hbox{$\sim$}}\; }
\def\gsim{\; \raise0.3ex\hbox{$>$\kern-0.75em
      \raise-1.1ex\hbox{$\sim$}}\; }
\nc{\nn}{\nonumber \\*}
\nc{\eq}[1]{\mbox{Eq.~(\ref{#1})}}

\nc{\annp}[3]{{\it  Ann.\ Phys.\ (N.Y.)\ }{{\bf #1} {(#2)} {#3}}}
\nc{\apl}[3]{{\it  Appl. Phys. Lett. }{{\bf #1} {(#2)} {#3}}}
\nc{\apj}[3]{{\it  Ap.\ J.\ }{{\bf #1} {(#2)} {#3}}}
\nc{\apjl}[3]{{\it  Ap.\ J.\ Lett.\ }{{\bf #1} {(#2)} {#3}}}
\nc{\app}[3]{{\it Astropart.\ Phys.\ }{{\bf #1} {(#2)} {#3}}}
\nc{\cmp}[3]{{\it  Comm.\ Math.\ Phys.\ }{{ \bf #1} {(#2)} {#3}}}
\nc{\cqg}[3]{{\it  Class.\ Quant.\ Grav.\ }{{\bf #1} {(#2)} {#3}}}
\nc{\epl}[3]{{\it  Europhys.\ Lett.\ }{{\bf #1} {(#2)} {#3}}}
\nc{\ijmp}[3]{{\it Int.\ J.\ Mod.\ Phys.\ }{{\bf #1} {(#2)} {#3}}}
\nc{\ijtp}[3]{{\it Int.\ J.\ Theor.\ Phys.\ }{{\bf #1} {(#2)} {#3}}}
\nc{\jmp}[3]{{\it  J.\ Math.\ Phys.\ }{{ \bf #1} {(#2)} {#3}}}
\nc{\jpa}[3]{{\it  J.\ Phys.\ A\ }{{\bf #1} {(#2)} {#3}}}
\nc{\jpc}[3]{{\it  J.\ Phys.\ C\ }{{\bf #1} {(#2)} {#3}}}
\nc{\jap}[3]{{\it J.\ Appl.\ Phys.\ }{{\bf #1} {(#2)} {#3}}}
\nc{\jpsj}[3]{{\it J.\ Phys.\ Soc.\ Japan\ }{{\bf #1} {(#2)} {#3}}}
\nc{\lmp}[3]{{\it Lett.\ Math.\ Phys.\ }{{\bf #1} {(#2)} {#3}}}
\nc{\mpl}[3]{{\it  Mod.\ Phys.\ Lett.\ }{{\bf #1} {(#2)} {#3}}}
\nc{\ncim}[3]{{\it  Nuov.\ Cim.\ }{{\bf #1} {(#2)} {#3}}}
\nc{\np}[3]{{\it  Nucl.\ Phys.\ }{{\bf #1} {(#2)} {#3}}}
\nc{\pr}[3]{{\it Phys.\ Rev.\ }{{\bf #1} {(#2)} {#3}}}
\nc{\pra}[3]{{\it  Phys.\ Rev.\ A\ }{{\bf #1} {(#2)} {#3}}}
\nc{\prb}[3]{{\it  Phys.\ Rev.\ B\ }{{{\bf #1} {(#2)} {#3}}}}
\nc{\prc}[3]{{\it  Phys.\ Rev.\ C\ }{{\bf #1} {(#2)} {#3}}}
\nc{\prd}[3]{{\it  Phys.\ Rev.\ D\ }{{\bf #1} {(#2)} {#3}}}
\nc{\prl}[3]{{\it Phys.\ Rev.\ Lett.\ }{{\bf #1} {(#2)} {#3}}}
\nc{\pl}[3]{{\it  Phys.\ Lett.\ }{{\bf #1} {(#2)} {#3}}}
\nc{\prep}[3]{{\it Phys.\ Rep.\ }{{\bf #1} {(#2)} {#3}}}
\nc{\prsl}[3]{{\it Proc.\ R.\ Soc.\ London\ }{{\bf #1} {(#2)} {#3}}}
\nc{\ptp}[3]{{\it  Prog.\ Theor.\ Phys.\ }{{\bf #1} {(#2)} {#3}}}
\nc{\ptps}[3]{{\it  Prog\ Theor.\ Phys.\ suppl.\ }{{\bf #1} {(#2)} {#3}}}
\nc{\physa}[3]{{\it  Physica\ A\ }{{\bf #1} {(#2)} {#3}}}
\nc{\physb}[3]{{\it  Physica\ B\ }{{\bf #1} {(#2)} {#3}}}
\nc{\phys}[3]{{\it Physica\ }{{\bf #1} {(#2)} {#3}}}
\nc{\rmp}[3]{{\it  Rev.\ Mod.\ Phys.\ }{{\bf #1} {(#2)} {#3}}}
\nc{\rpp}[3]{{\it Rep.\ Prog.\ Phys.\ }{{\bf #1} {(#2)} {#3}}}
\nc{\sjnp}[3]{{\it Sov.\ J.\ Nucl.\ Phys.\ }{{\bf #1} {(#2)} {#3}}}
\nc{\spjetp}[3]{{\it Sov.\ Phys.\ JETP\ }{{\bf #1} {(#2)} {#3}}}
\nc{\yf}[3]{{\it Yad.\ Fiz.\ }{{\bf #1} {(#2)} {#3}}}
\nc{\zetp}[3]{{\it Zh.\ Eksp.\ Teor.\ Fiz.\  }{{\bf #1}  {(#2)} {#3}}}
\nc{\zp}[3]{{\it Z.\ Phys.\ }{{\bf #1} {(#2)} {#3}}}
\nc{\ibid}[3]{{\sl ibid.\ }{{\bf #1} {#2} {#3}}}
\def\be{\begin{equation}}
\def\ee{\end{equation}}
\def\bea{\begin{eqnarray}}
\def\eea{\end{eqnarray}}

\begin{document}
{\hfill HIP-1999-05/TH}
\vskip30pt
\title{TESTING B-BALL COSMOLOGY WITH THE CMB\footnote{Invited talk
at Strong and Electroweak Matter '98, 2-5 December 1998, Copenhagen, Denmark }}

\author{Kari Enqvist}

\address{Department of Physics, University of Helsinki and Helsinki Institute
of Physics\\
FIN-00014 University of Helsinki, Finland\\E-mail: Kari.Enqvist@helsinki.fi}


\maketitle\abstracts{In D-term inflation models, 
the fluctuations of squark fields in the flat directions
give rise to isocurvature 
density fluctuations stored in the Affleck-Dine condensate.
After the condensate breaks up in B-balls, 
these can be perturbations in the baryon number, or, 
in the case where the present neutralino density comes directly from 
B-ball decay, perturbations in the number of 
dark matter neutralinos. The latter case results in a large 
enhancement of the isocurvature 
perturbation. In this case, the requirement that the deviation 
of the adiabatic perturbations from scale invariance due to the 
Affleck-Dine field is not too large imposes a lower 
bound on the magnitude of the isocurvature fluctuation 
of about $10^{-2}$ times the 
adiabatic perturbation. This should be observable by MAP and PLANCK.  }

\section{AD condensate and B-ball decay}
The quantum fluctuations of the inflaton field give rise to
fluctuations of the energy density which are adiabatic \cite{eu}.
However, in the minimal supersymmetric standard model (MSSM),
or its extensions, the inflaton is not the only fluctuating field.
It is well known that the MSSM scalar field potential has
many flat directions \cite{drt}, along which a non-zero expectation value can 
form 
during inflation, leading to a condensate after
inflation, the so-called Affleck-Dine (AD) condensate \cite{ad}.
When the Hubble rate becomes becomes of the order of the curvature of
the potential, given by the susy breaking mass $m_S$, the condensate
starts to oscillate. At this stage
B-violating terms are comparable to the mass
term so that the condensate achieves a net baryonic charge. In the
AD baryogenesis scenario the
subsequent decay of the condensate will then generate the observable
baryon number \cite{toni}.

An important point is that
the AD condensate is not stable but typically breaks up into non-topological
solitons \cite{ks2,bbb1}
which carry baryon (and/or lepton) number \cite{cole2,ks1}
and are therefore called B-balls (L-balls). 
For baryogenesis considerations, the most promising direction is the
$d=6$ (``$u^cu^cu^c$'') direction \cite{bbb2}, on which we shall focus on in
the following.
The formation of the B-balls takes place with an efficiency $f_B$, likely
to be in the range 0.1 to 1.
The properties of the
B-balls depend on SUSY breaking and on the flat direction along which the
AD condensate forms. We will consider SUSY breaking mediated to the 
observable sector
by gravity. In this case the B-balls are unstable but long-lived,
decaying well after the electroweak phase transition has taken 
place \cite{bbb1}, with a natural order of 
magnitude for decay temperature $T_d\sim {\cal O}(1)\GeV$. 
This assumes a reheating temperature after inflation, $T_R$, 
is less than about $10^4$ GeV. 
Such a low value of $T_R$ is in fact 
necessary in D-term inflation models because 
the natural magnitude of the phase of the AD field, $\delta_{\rm CP}$, 
is of the order of 1 
in D-term inflation and along the d=6 direction
 AD baryogenesis implies that the baryon to entropy ratio is 
$\eta_B\sim \delta_{\rm CP} (T_R/10^9\GeV)$ \cite{bbbd}
so that $T_R\simeq {\cal O}(1)\GeV$ would be the most natural choice.
It is significant that a low reheating temperature
 can naturally be achieved in D-term inflation models, as these have 
discrete symmetries in order to ensure 
the flatness of the inflaton potential which can simultaneuously lead to
a suppression of the reheating temperature \cite{bbbd}.

\section{Fluctuations of the AD field}
The AD field 
$\Phi=\phi e^{i\theta}/\sqrt{2}\equiv (\phi_1+i\phi_2)/\sqrt{2}$ 
is a complex field and, in the currently favoured D-term inflation
models \cite{dti}, 
is effectively massless during inflation. Therefore both its
modulus and phase are subject to fluctuations with
\be
\delta \phi_i(\vec x)=\sqrt{V}\int {d^3k\over (2\pi)^3}e^{-i\vec k\cdot\vec x}
\delta_{\vec k}~~,
\ee
where $V$ is a normalizing volume and where the power spectrum is the same 
as for the inflaton field,
\be
{k^3\vert\delta_{\vec k}\vert^2\over 2\pi^2}=\left({H_I\over 2\pi}\right)^2~~,
\ee
where $H_I$ is the value of the Hubble parameter during inflation.

 In D-term inflation models 
the phase of the AD field receives no 
order $H$ corrections after inflation and so its fluctuations are 
unsuppressed \cite{kmr}. 
The fluctuations of the phase correspond to 
fluctuations in the local baryon number density, or isocurvature
fluctuations, while the fluctuations of the modulus give rise to adiabatic
density fluctuations. 
For given background values $\bar\theta$ and $\bar\phi$, 
(with $\bar\theta$ naturally of the order of 
1) one finds \cite{johncmb}
\be 
\left({\delta \theta\over\;Tan(\bar\theta)}\right)_k={H_I\over\;Tan(\bar\theta)\bar\phi}
={H_Ik^{-3/2}\over\sqrt{2}Tan(\bar\theta)\bar\phi_I}~~,
\ee
where $\phi_I$ is the value of $\phi$ when the perturbation leaves the
horizon. The magnitude of the AD field $\Phi$ remains at the non-zero 
minimum of its potential until 
$H\simeq m_S$, after which the baryon asymmetry
$n_B \propto Sin(\theta)$ forms. Thus the isocurvature fluctuation reads
\be
\left({\delta n_B\over n_B}\right)_k\equiv \delta^{(i)}_B
=\left({\delta \theta\over\; Tan(\bar\theta)}\right)_k
~~
\ee

The adiabatic fluctuations of the AD field
may dominate over the
inflaton fluctuations, with potentially adverse consequences 
for the scale invariance of the 
perturbation spectrum, thus imposing an upper 
bound on the amplitude of the AD field \cite{johncmb}. 
          In the simplest D-term inflation model,
the inflaton is coupled to the matter fields $\psi_-$ and $\psi_+$ 
carrying opposite  Fayet-Iliopoulos charges through a superpotential
term $W=\kappa S\psi_-\psi_+$ \cite{dti,kmr}. At one loop level the inflaton potential
reads
\be
V(S)=V_0+{g^4\xi^4\over 32\pi^2}\ln\left({\kappa^2S^2\over Q^2}\right)
\;\; ; \;\;\;\; V_0 = \frac{g^2 \xi^4}{2}       ~~,\ee
where $\xi$ is the Fayet-Iliopoulos term and $g$ the gauge coupling 
associated with it. COBE normalization fixes $\xi=6.6\times 10^{15}\GeV$. 
In addition, we must consider the contribution of the AD field to the 
adiabatic perturbation. During inflation, the potential of the 
$d=6$ flat AD field
is simply given by
\be
V(\phi)={\lambda^2\over 32 M^6}\phi^{10}~~.
\ee
Taking both $S$ and
$\phi$ to be slow rolling fields one finds that the 
adiabatic part of the invariant perturbation is given by \cite{johncmb}
\be
\zeta = \delta\rho/(\rho+p)=\frac 34{\delta \rho_{\gamma}^{(a)}\over \rho_{\gamma}}
\propto {V'(\phi)+V'(S)\over V'(\phi)^2+V'(S)^2} \delta \phi ~~.
\ee
Thus the field which dominates the spectral index of the perturbation will be that 
with the largest value of $V^{'}$ and $V^{''}$.

\section{A lower bound on the isocurvature amplitude}
The index of the power spectrum 
is given by $n=1+2\eta-6\epsilon$, 
where $\epsilon$ and $\eta$ are defined as
\be
\epsilon=\frac 12{M^2}\left({V'\over V}\right)^2~~,~~
\eta={M^2}{V''\over V}~~.
\ee
The present
lower bounds imply $|\Delta n| \lsim 0.2$. It is easy to find out that the
the condition that the spectral index is acceptably close to 
scale invariance essentially reduces to the condition that the spectral index is dominated 
by the inflaton, $V'(\phi)<V'(S)$ and $V''(\phi)<V''(S)$. 
The latter requirement turns out to be slightly more stringent and
implies a
lower bound on the AD condensate field \cite{johncmb} with 
$\phi\lsim 0.48 \left(g/\lambda\right)^{1/4}(M \xi)^{1/2}$.

As a consequence, there is
a lower bound on the isocurvature fluctuation amplitude. 
Because the B-ball is essentially a squark condensate, in 
R-parity conserving models its decay produces 
both baryons and neutralinos ($\chi$), which we assume to be the lightest
supersymmetric particles (LSPs),
with $n_{\chi}\simeq 3n_B$ \cite{bbb2,bbbdm}. This case is 
particularly interesting, as the simultaneous production of 
baryons and neutralinos may help to explain the 
remarkable similarity of the baryon and dark matter neutralino number 
densities \cite{bbb2,bbbdm}. 
With B-ball decay temperatures $T_d \sim {\cal O}(1)\GeV$,
the decay products no longer thermalize completely and, so long as $T_d$ is
low enough that they do not annihilate after B-ball decay \cite{bbbdm}, 
retain 
the form of the original AD isocurvature fluctuation. 
Therefore in this scenario the cold dark matter particles
can have both isocurvature and adiabatic density fluctuations, 
resulting in an enhancement of 
the isocurvature contribution relative to the baryonic case.

The total LSP number density is the sum of the thermal relic density
$n_\chi^{(th)}$ and the density $n_\chi^{(B)}=3f_Bn_B$
 originating from the B-ball decay.
(Their relative importance depends on $T_R$; for $T_R\lsim {\cal O}(1)
\GeV$ one would find $n_\chi^{(th)}\simeq 0$.)
The isocurvature fluctuation
imposed on the CMB photons is then found to be \cite{johncmb}
\be
{\delta\rho_\gamma^{(i)}\over \rho_\gamma}
\simeq -\frac 43\left(1+{m_B\over 3f_Bm_\chi}\right)
\left({\Omega_\chi-\Omega_\chi^{(th)}\over \Omega_m}\right)\delta^{(i)}_B
\equiv -\frac 43\omega\delta^{(i)}_B~~,
\ee
where $\rho_\chi^{(B)}$ is the LSP mass density from the B-ball, 
$\Omega_m~(\Omega_\chi)$ is total matter (LSP) 
density (in units of the critical density). 
Thus 
\be
\beta\equiv \left( {\delta\rho_\gamma^{(i)}\over 
\delta\rho_\gamma^{(a)}}\right)^2
= \frac 19 \omega^2
\left({M^2V'(S)\over V(S)Tan(\bar\theta)\bar\phi}\right)^2~~,
\ee
It then follows that the lower limit on $\beta$ is
\be
\beta\gsim 2.5 \times 10^{-2}g^{3/2}\lambda^{1/2}\omega^2
Tan(\bar\theta)^{-2}~~.
\ee
Thus significant isocurvature fluctuations are a definite 
prediction of the AD mechanism.

Isocurvature perturbations give rise
to extra power at large angular scales but are damped at small angular
scales \cite{stomper}. The 
amplitude of the rms mass fluctuations in an $8h^{-1}$ Mpc$^{-1}$
sphere, denoted as $\sigma_8$, is about an order of magnitude lower 
than in the adiabatic case. Hence COBE normalization alone is
sufficient to set a tight limit on the relative strength of
the isocurvature amplitude. Small isocurvature fluctuations
are, however, beneficial, in the sense that they would improve the
fit to the power spectrum in $\Omega_{0} = 1$ 
CDM models with a cosmological constant \cite{axion}
(or $\Omega_{0} = 1$, $\Lambda = 0$  
CDM models with some hot dark matter \cite{burns}). 

Detecting isocurvature fluctuations at the level
of $\beta\sim 10^{-4}$ should be quite realistic \cite{johncmb} at MAP and 
Planck. Thus the forthcoming CMB experiments offer a test not only
of the inflationary Universe but also of the B-ball variant of AD baryogenesis.

\section*{Acknowledgments}
I wish to thank John McDonald for many discussions and an enjoyable 
collaboration. This work has been supported by the
 Academy of Finland  under the contract 101-35224.

\end{document}